\documentclass{aa}  

\usepackage{graphicx}
\usepackage{hyperref}
\usepackage{bookmark}
\usepackage{hyperref}
\usepackage{txfonts}
\usepackage[T1]{fontenc}
\usepackage{graphicx}
\usepackage{xcolor}
\usepackage{amsmath}
\usepackage{verbatim}
\usepackage{float}
\usepackage{lmodern}
\usepackage{anyfontsize}
\usepackage{comment}
\usepackage[switch]{lineno}  
\usepackage{newtxtext,newtxmath}
\usepackage{wasysym}
\usepackage{orcidlink}
\usepackage[T1]{fontenc}
\usepackage{graphicx}
\usepackage{xcolor}
\usepackage{subfig}
\usepackage{subcaption}   
\begin{document}

   \title{An energetic absorption outflow in QSO J1402+2330: \\Analysis of DESI observations}


   \author{M. Dehghanian\inst{1}\fnmsep\thanks{\email{m.dehghanian@uky.edu}}\orcidlink{0000-0002-0964-7500}  
          \and
          N. Arav\inst{2}\orcidlink{0000-0003-2991-4618}
          \and
          M. Sharma\inst{2}\orcidlink{0009-0001-5990-5790}
          \and
          G. Walker\inst{2}\orcidlink{0000-0001-6421-2449}
          \and
          K. Johnston \inst{2}
          \and
          M. Kaupin \inst{2}}

  \institute{$^{1}$Department of Physics and Astronomy, The University of Kentucky, Lexington, KY 40506, USA\\
   $^{2}$Department of Physics, Virginia Tech, Blacksburg, VA 24061, USA\\ }
   \date{Received XX, YYYY; accepted XX, YYYY}

 
  \abstract
   {Quasar outflows play a significant role in the active galactic nucleus (AGN) feedback, impacting the interstellar medium and potentially influencing galaxy evolution. Characterizing these outflows is essential for understanding AGN-driven processes.}
   {We aim to analyze the physical properties of the mini-broad absorption line outflow in quasar J1402+2330 using data from the Dark Energy Spectroscopic Instrument (DESI) survey. We seek to measure the outflow's location, energetics, and potential impact on AGN feedback processes.}
   {In the spectrum of J1402+2330, we identify multiple ionic absorption lines, including ground and excited states. We measure the ionic column densities and then use photoionization models to determine the total hydrogen column density and ionization parameter of the outflow. We utilized the population ratio of the excited state to the ground state of \ion{N}{iii} and \ion{S}{iv} to determine the electron number density.}
   {The derived electron number density, combined with the ionization parameter, indicates an outflow distance of approximately 2200 pc from the central source. Having a mass outflow rate of more than one thousand solar masses per year and a kinetic energy output exceeding 5$\%$ of the Eddington luminosity, this outflow can significantly contribute to AGN feedback.}
   {Our findings suggest the absorption outflow in J1402+2330 plays a potentially significant role in AGN feedback processes. This study highlights the value of DESI data in exploring AGN feedback mechanisms.}


   \maketitle
%

\section{Introduction}

Quasar outflows are invoked as key contributors to active galactic nuclei (AGN) feedback (e.g., 
\citealt{silk98, scan04, yuan18, vayner21, he22}). Absorption outflows observed in rest-frame UV spectra are generally 
classified into broad absorption lines (BALs, $\geq 2000$ km s$^{-1}$), narrow absorption lines (NALs, $\leq 500$ km s$^{-1}$), and mini-BALs. 
Mini-BALs exhibit absorption features with intermediate widths between NALs and traditional BALs, with smooth BAL-like profiles but velocity widths $\leq$2000 km s$^{-1}$ \citep{wey91,chur99,ham04,itoh20,ham13}.
Mini-BAL trough are typically unblended, allowing for reliable determination of ionic column densities. As a result, mini-BALs serve as effective probes of the physical conditions within outflows\citep{gang08,deh24,deh24b}.

The Dark Energy Spectroscopic Instrument (DESI) is a cutting-edge, multi-object spectrograph designed to map the large-scale structure of the universe with remarkable precision. Installed on the Mayall 4-meter telescope at Kitt Peak National Observatory, DESI can observe over 5,000 targets simultaneously, enabling the measurement of redshifts for millions of galaxies and quasars across a vast area of the sky. As of June 2023, DESI has identified approximately 90,000 quasars in its Early Data Release \citep[EDR,][]{DESI}, representing about 2$\%$ of its planned survey, which aims to map over 35 million galaxies and quasars (From DESI lab\footnote{https://www.desi.lbl.gov/2023/06/13/the-desi-early-data-release-in-now-available/}). The EDR serves as a valuable resource for the scientific community to evaluate DESI’s data quality, calibration, and potential for preliminary cosmological analyses. 

Recently, some studies have utilized DESI data to detect/investigate the absorption lines. For instance, \cite{nap23} detected \ion{Mg}{ii} ($\lambda$ 2796.28\AA) absorbers via developing an autonomous supplementary spectral pipeline. This was followed by searching for other common metal lines, such as \ion{Fe}{ii}, \ion{C}{iv}, and \ion{Si}{iv}. In another study, \cite{fil24} present a catalog of BAL quasars identified in the DESI-EDR survey. The current paper marks the first instance of employing DESI data to analyze the energetics of an absorption outflow system. In this study, we investigate the quasar SDSS~J1402+2330, in which two distinct absorption outflow systems are detected. Using DESI spectra, we aim to: 
\begin{itemize}
    \item Measure the physical parameters of the low-velocity outflow, including ionic column densities ($N_{\textrm{ion}}$), total hydrogen column density ($N_{\textrm{H}}$), ionization parameter ($U_{\textrm{H}}$), electron number density ($n_{\textrm{e}}$), and distance from the AGN ($R$).
    \item Determine the outflow’s energetics, including the mass-loss-rate and kinetic luminosity, to assess its possible role in AGN feedback.
\end{itemize}

The paper's structure is as follows:
In Section~\ref{sec:obs}, we describe the observations and data acquisition of the target. Section~\ref{sec:anal}, details
the approaches used to calculate $N_{\textrm{ion}}$ of
the mini-BALs. From these $N_{\textrm{ion}}$, we derive the photoionization solution,  $n_{\textrm{e}}$ and $R$. We then calculate the black hole (BH) mass and the quasar’s
Eddington luminosity. These are followed by deriving the energetics of
the outflow system. Section~\ref{sec:disc} presents the summary of the paper and concludes it.

Here we adopt standard $\Lambda$CDM cosmology with h= 0.677, $\Omega_{m}$= 0.310, and
$\Omega_\Lambda$ = 0.690 \citep{plan20}. We used the Python astronomy
package Astropy \citep{astro13,astro18} for our
cosmological calculations, as well as Scipy \citep{virt20},
Numpy \citep{harr20}, and Pandas \citep{reba21} for most of our numerical computations. For our plotting purposes, we used Matplotlib \citep{hunt07}.
\section{Observation}
\label{sec:obs}
Quasar SDSS~J1402+2330 (hereafter J1402) is a 
bright quasar, at Redshift z=2.8302 (based on the NASA/IPAC Extragalactic 
Database)\footnote{NED:\url{https://ned.ipac.caltech.edu/}},
with J2000 coordinates at RA=14:2:21.52 and Dec=+23:30:43.32.
Observations of J1402 were conducted as part of the DESI-EDR survey over two nights: March 10, 2021, and April 4, 2021, with a total exposure time of 5769.5 seconds. The spectral data span the wavelength range of 3600 \AA\  to 9800 \AA, with a 
spectral resolution which falls between $R\approx$2000-5000, depending on the specific wavelength observed. The observations are part of the sv1 Survey Validation Phase and were processed with the DESI pipeline, and co-added spectra are referenced in file 3851 from the DESI archive. 

After obtaining the data using client developed by SPectra Analysis and Retrievable Catalog Lab (SPARCL), we identified two absorption line outflow systems from multiple ions (see figure~\ref{figFlux1}):
\begin{itemize}
    \item Low-velocity outflow: $v_{\textrm{centroid}}\approx-4300$ km~s$^{-1}$
    \item High-velocity outflow: $v_{\textrm{centroid}}\approx-8500$ km~s$^{-1}$
\end{itemize}
\begin{figure}
\includegraphics[width=\columnwidth]{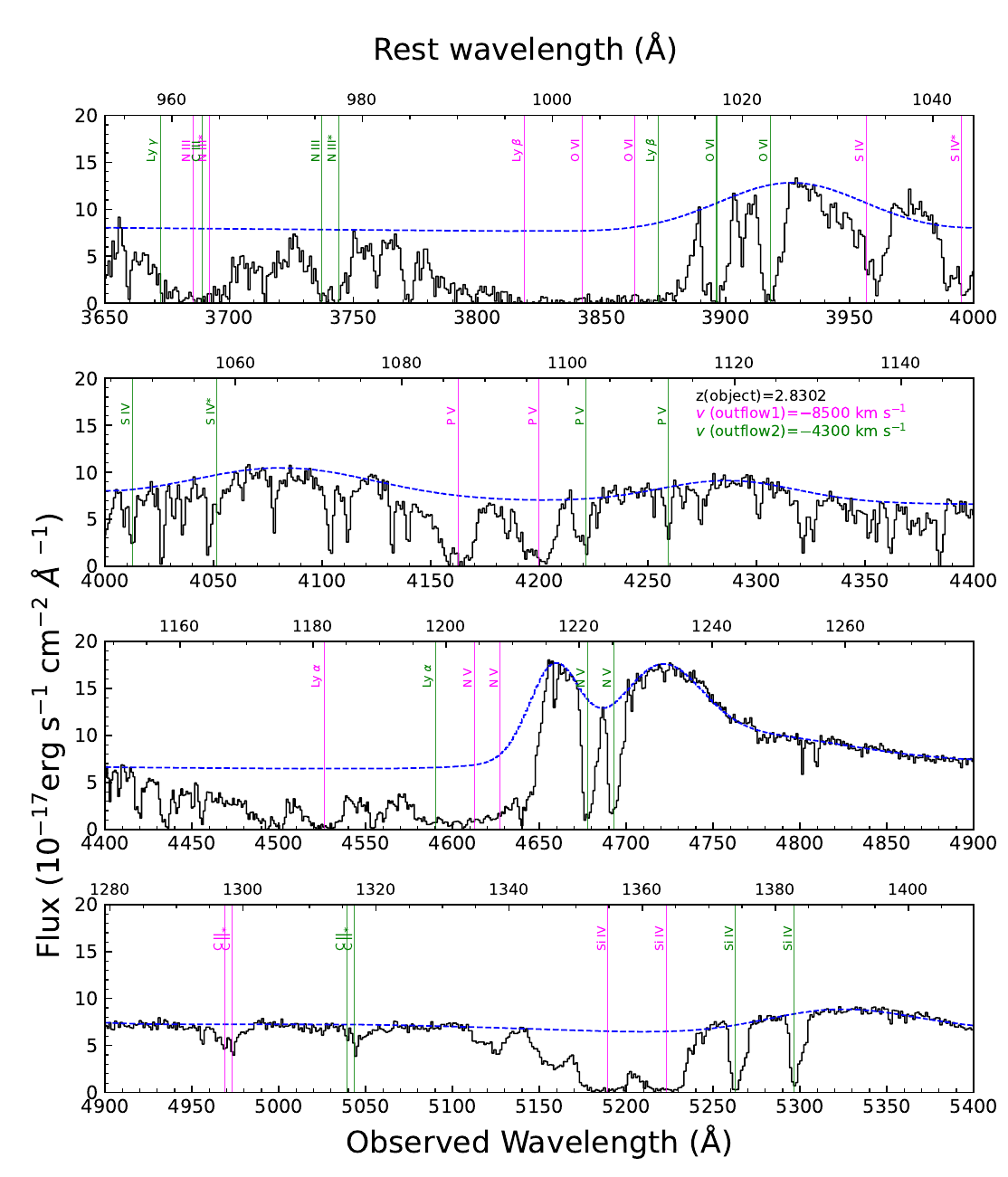}
\caption{The spectrum of SDSS J1402+2330 as observed by the DESI in 2021. The absorption features of the outflow system with a velocity of $-$8500 km s$^{-1}$ are marked with magenta lines, while the green lines show the absorption from the outflow system with a centroid velocity of $-$4300 km s$^{-1}$. The dashed blue line shows our continuum emission model.}
            \label{figFlux1}%
\end{figure}
\begin{figure}
\ContinuedFloat
\captionsetup{list=off}
\includegraphics[width=\columnwidth]{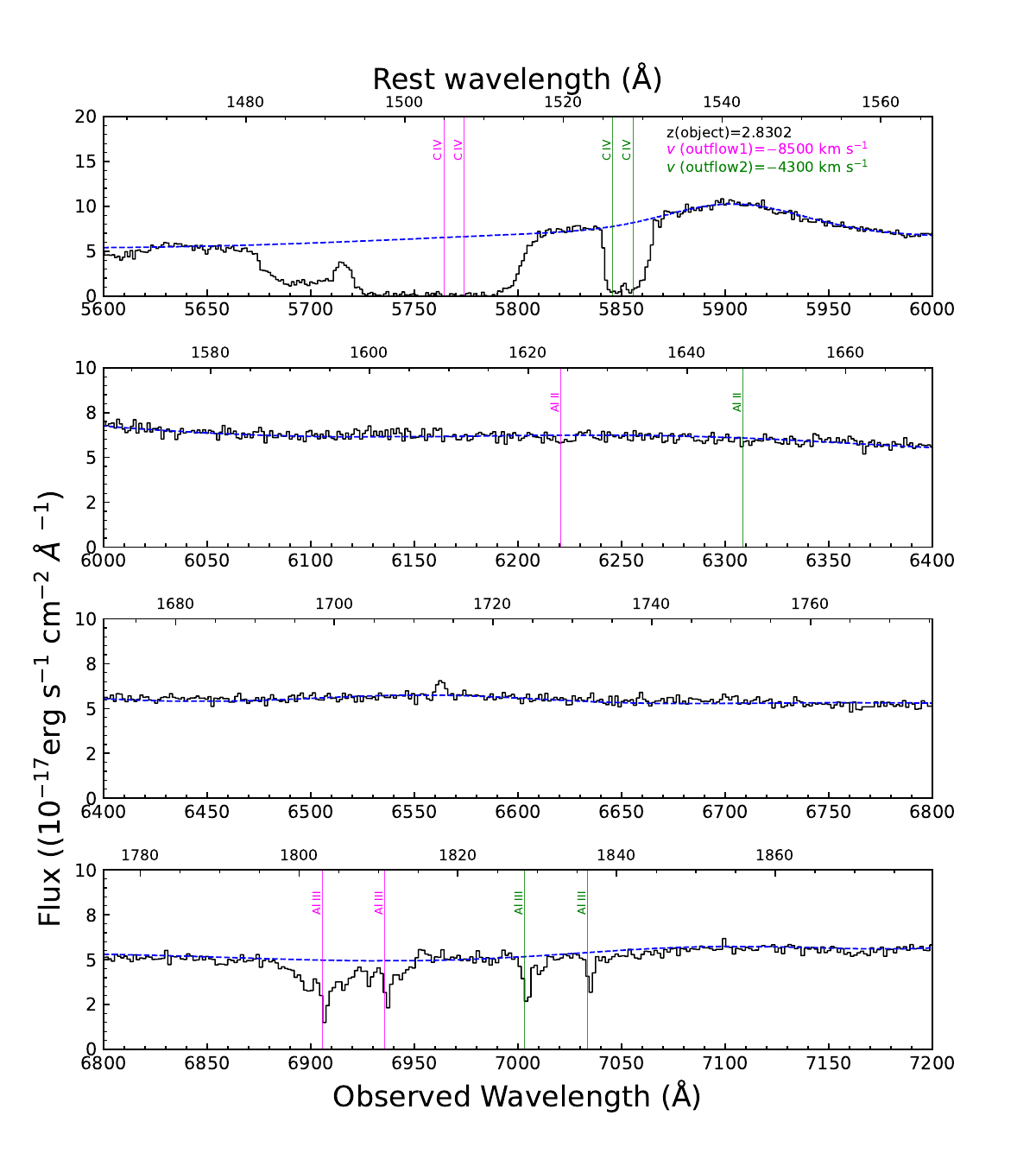}
\caption{Continued}
            \label{figFlux2}%
\end{figure}
The low-velocity outflow system is the focus of our analysis due to the detection of ground-state and excited-state absorption troughs in its spectrum, including \ion{N}{iii}, \ion{N}{iii*}, \ion{S}{iv}, \ion{S}{iv*}, \ion{C}{ii}, and \ion{C}{ii*} (see figure~\ref{figFlux1}). The presence of multiple ions in various energy states makes this outflow an especially valuable target for determining its energetics. This outflow has a centroid velocity of approximately $v_{\textrm{centroid}}\approx-4300$ km~s$^{-1}$
and exhibits absorption troughs from \ion{Al}{ii}, \ion{Al}{iii}, and \ion{P}{v} as well. The absorption width of this system is around 600 km~s$^{-1}$, which categorizes it as a mini-BAL. Throughout this paper, we refer to this high-velocity mini-BAL system as "the outflow." 

The high-velocity absorption outflow is excluded from analysis because reliable column density measurements cannot be obtained for the excited states identified in this system: Both \ion{S}{iv*} and \ion{N}{iii*} are saturated, and \ion{C}{ii*} is heavily blended, precluding a reliable assessment.
\section{Analysis}
\label{sec:anal}
Figures~\ref{figFlux1} displays the spectrum along with the identified absorption lines. As noted above, we focus on the low-velocity outflow system, for which the absorption lines are shown with vertical blue lines (high-redshift). We exclude \ion{O}{vi} doublet ($\lambda\lambda$ 1031.91,1037.61\AA), \ion{N}{v} doublet ($\lambda\lambda$ 1238.82, 1242.80\AA), and also \ion{Si}{iv} doublet ($\lambda\lambda$ 1393.76, 1402.77\AA) from our analysis, as their flux values approach zero, indicating saturation. Note that the spectrum has a velocity resolution ranging between 60 to 150 km~s$^{-1}$, with the velocity resolution around $\sim 150$ km~s$^{-1}$ at the \ion{N}{iii} region.

\begin{figure}
\includegraphics[width=\columnwidth]{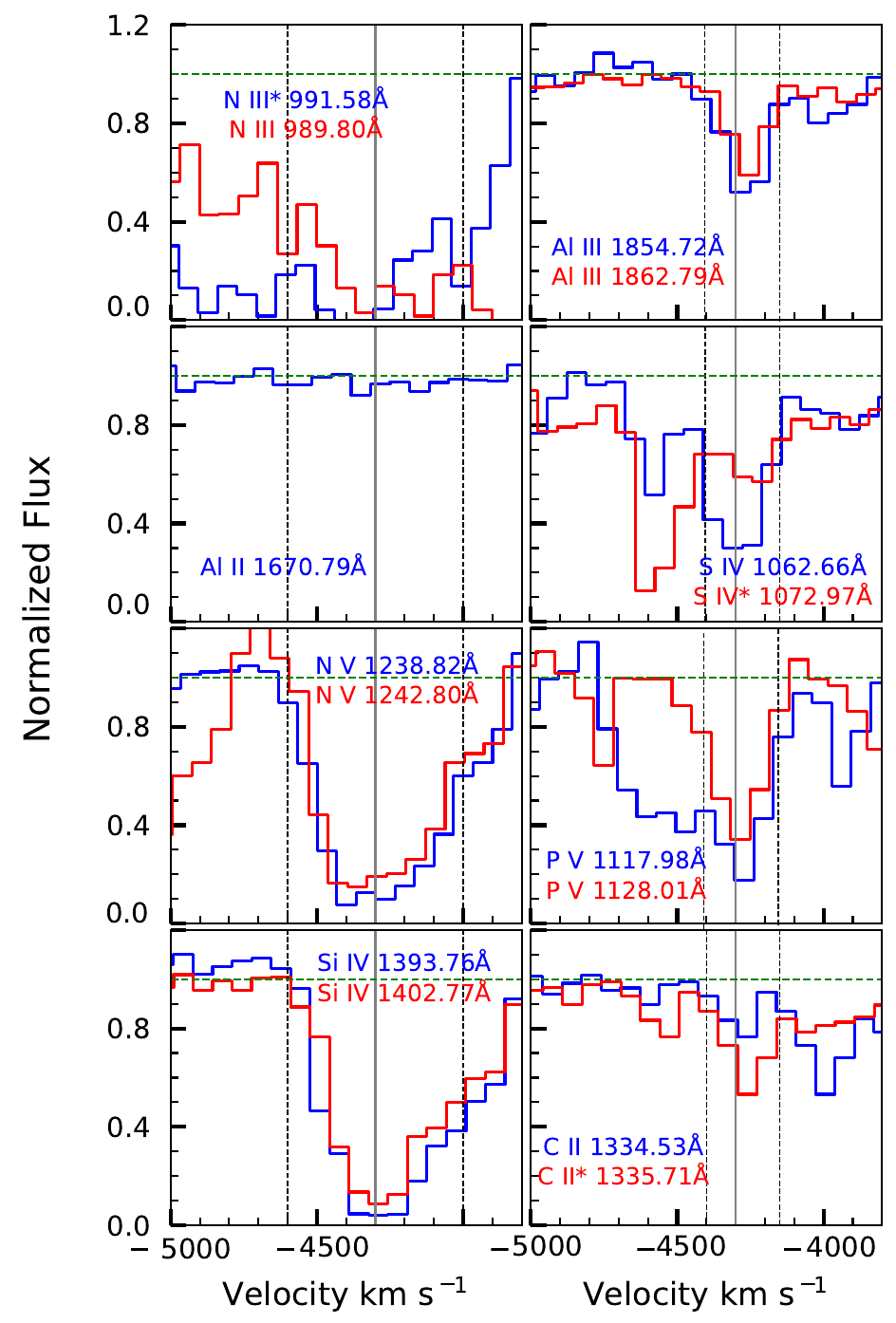}
\caption{Normalized flux versus velocity for outflow's absorption troughs detected in the spectrum of J1402. The horizontal green dashed line shows the continuum level,
and the vertical black dashed lines show the integration range (see text). The vertical solid gray line 
indicates the centroid velocity.}
            \label{velocity}%
\end{figure}

To identify the troughs in the spectrum and determine their ionic column densities, we first obtain a normalized spectrum
by dividing the data by the unabsorbed emission model. In the spectrum of J1402, we modeled the unabsorbed emission model with a single power law ($F_{\lambda}=6.2E$-$17(\frac{\lambda_{\textrm{rest}}}{\lambda_{0}})^{-0.521}$ with $\lambda_{0}=1350$ \AA), while emission regions are modeled with one or more Gaussian functions.
Figure \ref{velocity} presents the normalized flux versus velocity for some of the blue-shifted absorption lines detected in the spectrum of J1402. As shown in the figure, for most of the ions we integrate over a velocity range of $-4600$ to $-4000$ km~s$^{-1}$ (width=600 km s$^{-1}$). However, for \ion{S}{iv}, \ion{S}{iv*}, \ion{C}{ii}, \ion{C}{ii*}, and \ion{Al}{iii} the integration range is adjusted to $-4400$ to $-4150$ km~s$^{-1}$
to account for differences in absorption features: the regions outside this narrower range deviate from the profile of other absorption troughs and are therefore deemed unreliable for calculations. For \ion{P}{v}, the blue trough is excluded due to contamination by the Ly$\alpha$ forest, and the integration range for the red trough is similarly constrained to $-4400$ to $-4150$ km s$^{-1}$.
\subsection{Determining the ionic column densities}
\begin{figure}
\includegraphics[width=\columnwidth]{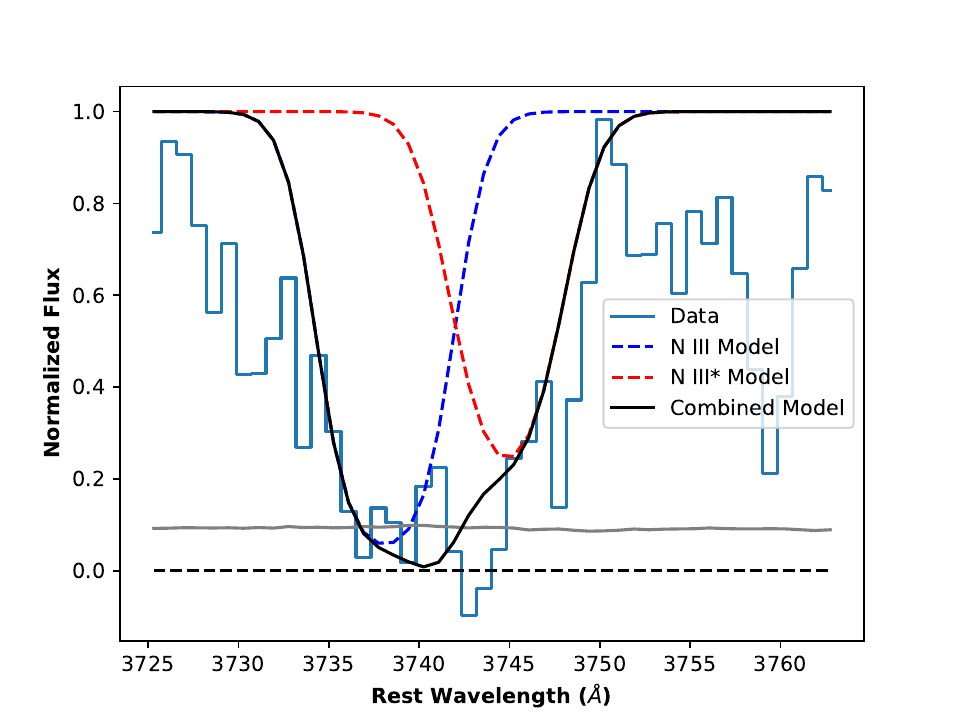}
\caption{Gaussian modeling of the \ion{N}{iii*}$ \lambda$991.58 \AA\ (blue dashed line) and \ion{N}{iii} $\lambda$989.80
\AA\ (red dashed line) absorption troughs. The black solid line shows the final model, which results from combining two Gaussian curves. The grey line shows the level of noise around the modeled region. }
            \label{figN3}
\end{figure}

To understand the physical characteristics of the outflow, we start by determining the individual $N_{\textrm{ion}}$ of the absorption troughs. One approach to determine the ionic column densities is the "apparent optical depth" (AOD) method, which assumes that the source is uniformly and completely covered by the outflow \citep{spit68,sava91}. In this scenario, the column densities are determined using equations(~\ref{eq-1}) and (\ref{eq0}) below \citep[e.g.,][]{spit68, sava91,arav01}:
\begin{equation}
I(v)=e^{-\tau(v)}, \label{eq-1}
\end{equation}
\begin{equation}
N_{\textrm{ion}}=\frac{3.77\times 10^{14}}{\lambda_{0}f}\times \int \tau(v)~dv 
 [\textrm{cm}^{-2}],
\label{eq0}
\end{equation}
\noindent where $I(v)$ is the normalized intensity profile as a function of velocity and $\tau(v)$ is the optical depth of the absorption trough. $\lambda_{0}$ refers to the transition's wavelength, $f$ is the oscillator strength of the transition, and and $v$ is measured in km~s$^{-1}$.
To determine $N_{\textrm{ion}}$ of the \ion{S}{iv} and \ion{S}{iv*}, we opted for the AOD technique, as they arise from different energy levels and are not saturated, since they are much shallower than the saturated troughs like \ion{O}{vi}. Since the absorption troughs of these ions are not very deep, these can be assumed to be unsaturated, and their column density measurements via the AOD method are reliable. For \ion{P}{v}, we focus exclusively on its red trough ($\lambda$1128.00\AA) and treat it as a singlet. This is because the blue trough is clearly contaminated by Ly$\alpha$ forest absorption (see figure~\ref{velocity}).  
We are unable to directly apply the AOD method to \ion{N}{iii} and \ion{N}{iii*} absorption troughs, since they are blended (see figure~\ref{velocity}). Instead, we model each ion with a Gaussian function in flux space, ensuring both Gaussians share the same centroid velocity ($v_{\textrm{centroid}}$) and width (see figure~\ref{figN3}). We then apply the AOD method to each Gaussian individually and sum the results to determine the final \ion{N}{iii} ionic column density.
For more details regarding the AOD method, see \cite{arav01}, \cite{gabel03}, and \cite{byun22c}. 

\begin{figure}  
    \centering
    \begin{minipage}{\linewidth}  
        \centering
        \includegraphics[width=1\linewidth]{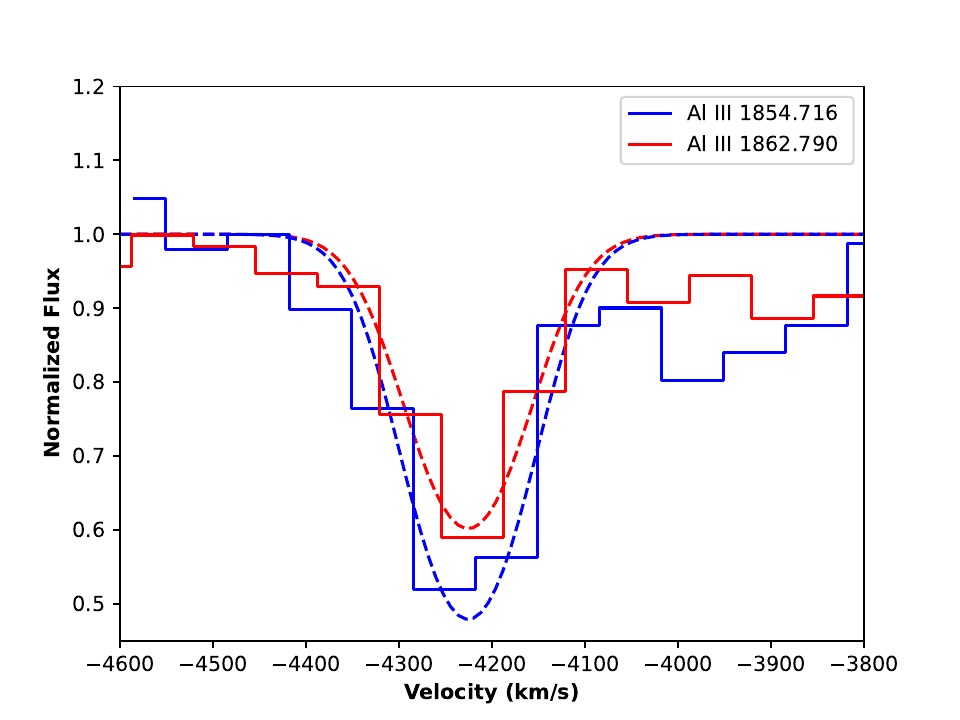}
        \vspace{-0.76 cm}
    \end{minipage}
    \caption{Gaussian modeling of the \ion{Al}{iii} $\lambda$1854.72 \AA\ (blue) and the \ion{Al}{iii} $\lambda$1862.79 (red) absorption troughs. In both cases, the data are shown in histograms while the fits are shown with dashed lines of the same color.}
    \label{figAl3}
\end{figure}

The partial covering (PC) method is applied when two or more lines originate from the same energy level. This method assumes that the outflow partially covers a homogeneous source\citep{barl97, arav99a, arav99b}. In this method, since a velocity-dependent covering factor is determined, the effects of non-black saturation \citep{kool02} are considered. To use the PC method, we calculate the covering fraction $C(v)$ and optical depth $\tau (v)$ using equations~(\ref{eq30}) and (\ref{eq40}) \citep[see][]{arav05}. For doublet transitions, where the blue component has twice the oscillator strength of the red component:
\begin{eqnarray}
    I_R(v)-[1-C(v)]=C(v)e^{-\tau(v)} \label{eq30}
\end{eqnarray}
and
\begin{eqnarray}
    I_B(v)-[1-C(v)]=C(v)e^{-2\tau(v)}\label{eq40}
\end{eqnarray}
In the equations above, $I_R(v)$ and $I_B(v)$ are the normalized intensities of the red and blue absorption features, respectively. $\tau(v)$ is the optical depth profile of the red component. In this study, we apply the PC method to the \ion{Al}{iii} doublet ($\lambda\lambda$ 1854.71, 1862.78\AA) since they are both originating from the ground state. 
We model each trough with a Gaussian in flux space, where both Gaussian functions have the same centroid velocity ($v_{\textrm{centroid}}$) and width (see figure~\ref{figAl3}). This enables us to calculate the values of $\tau(v)$ and $C(v)$. We then use the resultant values in the PC method equations to calculate the column density of \ion{Al}{iii}.
See \cite{barl97,arav99a,arav99b, kool02, arav05,borg12b,byun22b, byun22c} to find detailed explanations of both methods.

We also consider the systematic errors arising from the choice of the continuum (See figure 4 of \citep{deh24b}). To include this source of error in our calculations, we assume a 10$\%$ systematic error. The final adopted errors for ionic column densities are determined by quadratically combining the AOD errors with the 10$\%$ systematic uncertainty in the local continuum level. For \ion{N}{iii} and \ion{Al}{iii}, the errors derived from the Gaussian fits are quadratically added to the assumed systematic error.  Table~\ref{tab1} summarizes the final results of our $ N_{\textrm{ion}}$ measurements,
along with their associated uncertainties. In this table, two ions are treated as upper limits rather than accurate measurements: \ion{S}{iv*}, due to contamination by the Ly$\alpha$ forest, and \ion{Al}{ii}, as its absorption trough is almost undetectable (see figures~\ref{figFlux2} and \ref{velocity}).

\begin{table}[ht!]
\caption{\label{tab1}Ionic column densities}
\centering
\begin{tabular}{lccc}
\hline\hline
Ion&$N_{\textrm{ion}}$&Upper Uncertainty&Lower Uncertainty\\
\hline
\ion{Al}{ii}&<2.7&1.0& -- \\
\ion{Al}{iii}&77&11&$-$11 \\
\ion{P}{v} &290 &37&$-$35\\
\ion{S}{iv}&1500&160&	$-$120\\
\ion{S}{iv*}&<740&	76&--\\
\ion{S}{iv} (total)$^{a}$&2300 &290&$-$700\\
\ion{N}{iii}&4000&900&$-$900\\
\ion{N}{iii*}&2200&400&$-$400\\
\ion{N}{iii} (total)&6200&1200&$-$1200\\
\ion{C}{ii}&77&12&11\\
\ion{C}{ii*}&170&15&14\\
\ion{C}{ii} (total)&240&31&$-$30\\
\hline
\end{tabular}
\tablefoot{The $N_{\textrm{ion}}$ of the absorption lines detected in the J1402 outflow. All of the column density 
values are in units of 10$^{12}$ cm$^{-2}$.\\ 
$^{a}$The $N_{\textrm{ion}}$ for \ion{S}{iv} (total) is the sum of ionic column densities of \ion{S}{iv} 1062.66\AA\  and \ion{S}{iv*} 1072.97\AA. To determine the total uncertainties, we quadratically add the errors in the column densities together with a 10$\%$ systematic error.
The same logic works for \ion{N}{iii} (total), and \ion{C}{ii} (total). For our photoionization modeling, the "total" values will be used.}
\end{table}

\subsection{Photoionization modeling}
\label{sec:photo}
The gas in quasar outflows is in photoionization equilibrium, where the balance between ionization and recombination processes determines the ionization state of the medium \citep[and references therein]{dav79,krol99, oster06}. Photoionization modeling provides a framework for interpreting the $N_{\textrm{ion}}$ observed in the outflowing gas, allowing us to constrain its physical properties. To achieve this, we use the spectral synthesis code Cloudy \citep{cloudy23}, in combination with the measured $N_{\textrm{ion}}$ to estimate $N_{\textrm{H}}$ and $U_{\textrm{H}}$ of the outflow. 
Cloudy solves the equations of photoionization equilibrium in the outflow, which we model as a plane-parallel slab with constant hydrogen number density and abundances, ionized by a specified spectral energy distribution (SED). 
In this approach, we use Cloudy to produce photoionization models that predict all $N_{\textrm{ion}}$ values for a grid of $N_{\textrm{H}}$ and $U_{\textrm{H}}$ values. We then compare these produced $N_{\textrm{ion}}$ to the observed ones to find the $N_{\textrm{H}}$ and $U_{\textrm{H}}$ values that best reproduce the observed $N_{\textrm{ion}}$.
This methodology is the same as was used in previous studies \citep[e.g.,][]{byun22a, byun22b, byun22c, walk22, deh24,deh24b, mayank24}.

\begin{figure}  
    \centering
    \begin{minipage}{\linewidth}  
        \centering
        \includegraphics[width=1\linewidth]{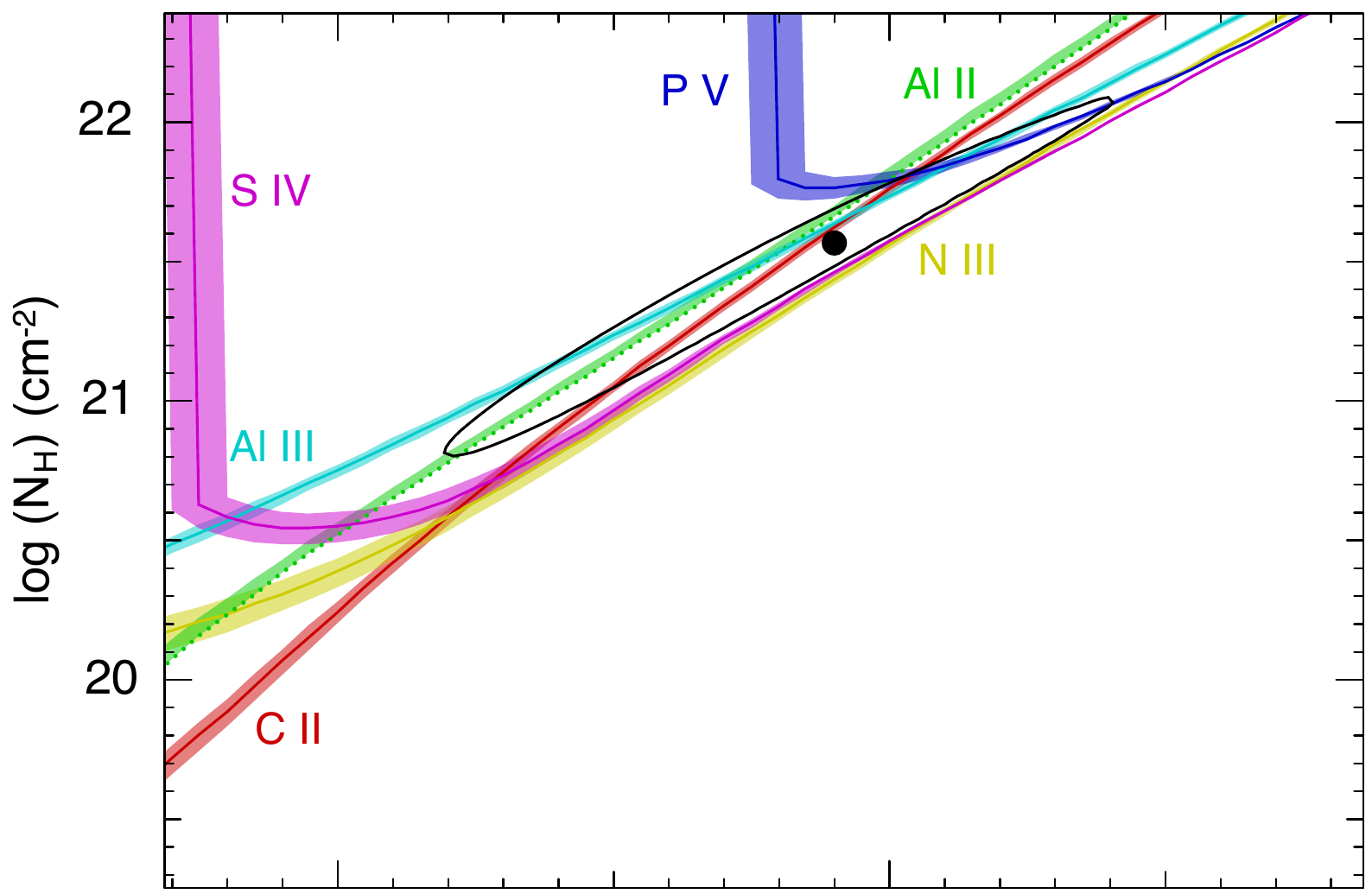}
        \vspace{-0.20 cm}
    \end{minipage}
    \begin{minipage}{\linewidth}  
        \centering
        \includegraphics[width=1\linewidth]{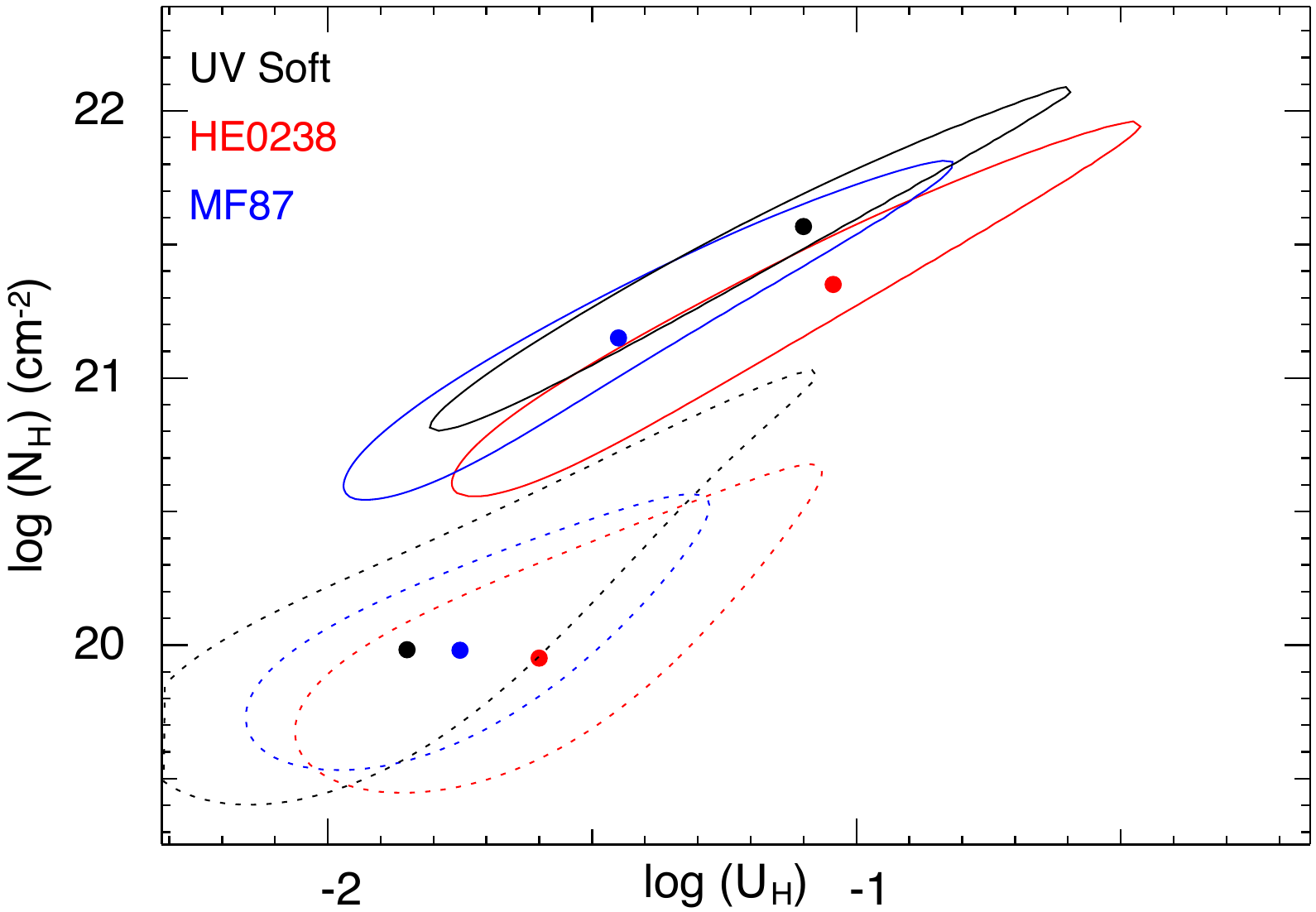}
    \end{minipage}

    \caption{Top: Phase plot showing the Photoionization solution for the absorption low-velocity outflow system in J1402, using the UV-soft SED and solar abundances. Each solid line represents the range of models ($U_{\textrm{H}}$ and $N_{\textrm{H}}$) that predict a column density matching the observed value for that ion. The shaded bands are the uncertainties associated with each $N_{\textrm{ion}}$ measurement. The dotted line (\ion{Al}{ii}) represents an upper limit. The black dot shows the solution, which is surrounded by $\chi^{2}$, as the black oval.\\ 
    Bottom: The photoionization solution for a total of six models, including three SEDs (HE0238, MF87, and UV–soft) and two abundances: solar (shown with solid lines) and super-solar (shown with dashed lines)}.
    \label{NVUapp}
\end{figure}

Note that the photoionization state of an outflow depends on the assumed abundances and the SED incident upon it. To consider these effects, we explored three SEDs, including HE0238 \citep{arav13}, MF87 \citep{mat87}, and UV-Soft SED \citep{dun10}, and two sets of abundances: solar ($Z=Z_{\astrosun}$) and super-solar metallicity (Z=4.68 $\times$ Z$_{\astrosun}$) \citep{ball08}, a total of six models  (See figure 5 of \cite{deh24b} for a visual comparison of the three SEDs. The accompanying discussion in the text provides further insights into the characteristics of these SEDs). The hydrogen column densities, ionization parameters, and the reduced $\chi^{2}$ resulting from these models are presented in  Table~\ref{tab2}.
\begin{table}[ht!]
\caption{\label{tab2}Photoionization solution for six models}
\centering
\begin{tabular}{lcccc}
\hline\hline
SED&metalicity&$\textrm{log}(N_{\textrm{H}}) [\textrm{cm}^{-2}]$&$\textrm{log}(U_{\textrm{H}})$&$\chi^{2}_{red}$\\
\hline
HE0238&solar&$21.3^{+0.6}_{-0.8}$ & $-1.0^{+0.6}_{-0.7}$&57\\
\\

MF87&solar& $21.1^{+0.7}_{-0.6}$ &$-1.5^{+0.6}_{-0.5}$&52\\
\\
UV-soft &solar&$21.6^{+0.5}_{-0.8}$ &$-1.1^{+0.5}_{-0.7}$&50\\
\\
HE0238&super-solar&$19.9^{+0.7}_{-0.5}$ & $-1.6^{+0.5}_{-0.5}$&61\\
\\
MF87&super-solar&$20.0^{+0.6}_{-0.4}$ & $-1.7^{+0.5}_{-0.4}$&50\\
\\
UV-soft&super-solar&$20.0^{+1}_{-0.6}$ & $-1.8^{+0.8}_{-0.5}$&72\\
\hline
\end{tabular}
\end{table}
We select the UV-soft SED with solar metalicity as our preferred model since this SED is more appropriate to be used for high-luminosity quasars \citep{dun10}, compared to MF87 (which has the same $\chi^{2}_{red}$ when paired with super-solar metalicity). Figure~\ref{NVUapp}, upper panel, shows the photoionization modeling using the UV-soft SED and solar metallicity. In this figure, the solid lines show the $N_{\textrm{ion}}$ taken as measurements, while shaded bands are the uncertainties associated with these measurements (see Table~\ref{tab1}). The green dotted line indicates that the ionic column density of that specific ion (\ion{Al}{ii}) is considered to be an upper limit. The black dot shows the best $\chi^{2}$-minimization solution. The lower panel of Figure~\ref{NVUapp} compares all six models. In this panel, the dots surrounded by the solid-line ovals  show the best solution when using each SED and solar metallicity, while the dots surrounded by the dashed-line ovals  show the best solution when using each
SED and super-solar metallicity.

\subsection{Electron number density}
\label{sec:ne}
\begin{figure}
\includegraphics[width=\columnwidth]{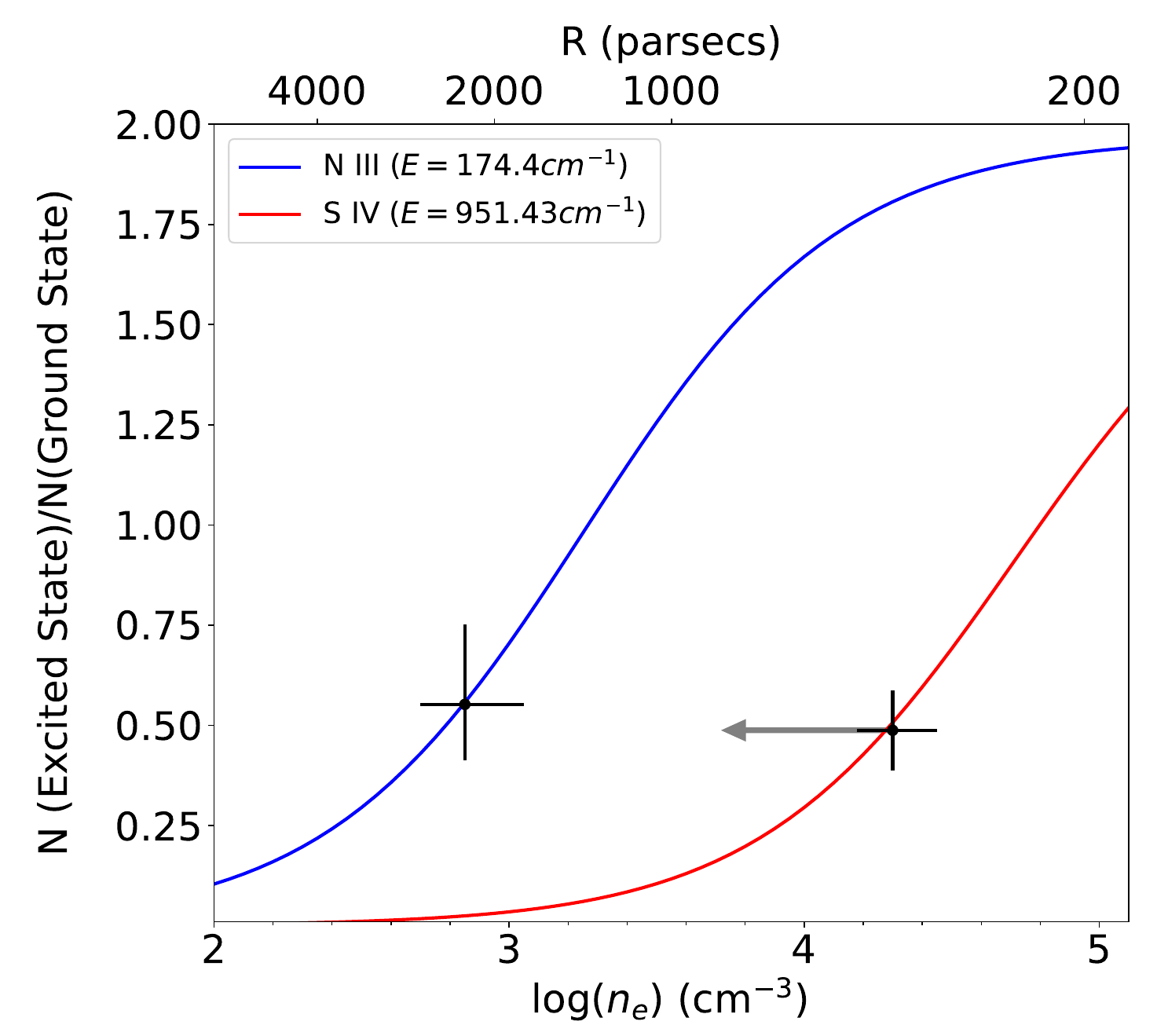}
      \caption{ Density diagnostic for the low-velocity absorption outflow. The measured $\frac{N_{\text{\ion{N}{iii*}}}}{N_{\text{\ion{N}{iii}}}}$ and $\frac{N_{\text{\ion{S}{iv*}}}}{N_{\text{\ion{S}{iv}}}}$ are overlayed on their theoretical level population curves, blue and red respectively, which are obtained for a temperature of $T=15000$ K.}
         \label{figchi}
\end{figure}

Assuming collisional excitation, we can determine $n_{\textrm{e}}$ of the outflow \citep[see][]{arav18} using the population ratio of excited and ground states of the \ion{N}{iii} ion. To do so, we use the Chianti 9.0.1 atomic database \citep{dere97, dere19} to predict the abundance ratios of the excited state to the resonance state for \ion{N}{iii} as a function of n$_{e}$. These calculations are affected by the temperature of the outflow. We assume a temperature of 15,000 K, predicted by Cloudy for our best $U_{\textrm{H}}$/$N_{\textrm{H}}$ solution (see Table~\ref{tab2}). The blue line in Figure~\ref{figchi} displays the predicted ratio versus $n_{\textrm{e}}$ for \ion{N}{iii}, while the red line shows the same for \ion{S}{iv}. The black dot in both cases has a y-value equal to our measurements of the $N_{\textrm{ion}}$ ratio, resulting in an estimation for $n_{\textrm{e}}$ (x-value). Note that while we accept $\frac{\ion{N}{iii*}}{\ion{N}{iii}}$ ratio as a measurement, $\frac{\ion{S}{iv*}}{\ion{S}{iv}}$ is an upper limit. This is because \ion{S}{iv*} is possibly contaminated by the Ly$\alpha$ forest, making it an upper limit rather than an accurate measurement. 

To estimate the uncertainties on $n_{\textrm{e}}$, we first calculate the uncertainties in the column density ratio, $\frac{N_{\text{\ion{N}{iii*}}}}{N_{\text{\ion{N}{iii}}}} = 0.55^{+0.16}_{-0.1}$ \citep[for detailed explanations see][]{deh24b}. These uncertainties are represented by vertical black lines. Using these values, we derived the corresponding errors in $n_{\textrm{e}}$, which are shown by horizontal black lines. A similar approach was applied for \ion{S}{iv}; however, instead of a lower bound for the electron number density, we included an arrow to indicate that this ratio is indeed an upper limit and the true $n_{\textrm{e}}$ could be any value below this threshold. Therefore, we base our measurement on the $\frac{\ion{N}{iii*}}{\ion{N}{iii}}$ ratio that yields $\text{log}(n_{e})= 2.85^{+0.20}_{0.15}$ [cm$^{-3}$], which is consistent with the upper limit acquired from $\frac{\ion{S}{iv*}}{\ion{S}{iv}}$ ratio ($\text{log}(n_{e})\leq 4.3$ [cm$^{-3}$], see Fig.~\ref{figchi}). 

The top axis of Figure~\ref{figchi} displays the distance between the outflow and the central source in parsecs. Below, we describe the method used to convert $n_{\textrm{e}}$ into distance.

\subsection{Distance of the outflow from the AGN}
We can now use the definition of $U_{\textrm{H}}$ (equation (14.4) from \cite{oster06}) to calculate the distance between the outflow and the AGN($R$):
\begin{equation}
U_H\equiv\frac{Q(\textrm{H})}{4\pi R^{2}c~n_{\textrm{H}}} \Rightarrow R=\sqrt{\frac{Q(\textrm{H})}{4\pi c~n_{\textrm{H}} U_{\textrm{H}}}} \label{eq1}
\end{equation}

\noindent $U_{\textrm{H}}$ is already determined from the photoionization modeling (see Section~\ref{sec:photo}), $n_{\textrm{H}}$ is the hydrogen density which for a highly ionized plasma is estimated to be  $n_{\textrm{H}}\approx \frac{n_{\textrm{e}}}{1.2}$ \citep{oster06}. c
is the speed of light, and $Q(\textrm{H})$ (s$^{-1}$) is the number of hydrogen-ionizing photons emitted by the central object per second. To determine $Q(\textrm{H})$, we scale the UV-soft SED with the observed continuum flux of J1402 at $\lambda_{\textrm{observed}}$=5010 \AA. We use the quasar's redshift, adopted \(\Lambda\)CDM cosmology, and the scaled SED to obtain the bolometric luminosity and $Q(\textrm{H})$  \citep [see][]{byun22a,byun22b,walk22}:

Following these steps, we obtain $Q(\textrm{H})$=$8.12 \pm 0.23\times 10^{56}$ s$^{-1}$ and $L_{\textrm{bol}}$=$1.20\pm 0.05\times 10^{47}$ erg s$^{-1}$. Using these values, along with $n_{\textrm{H}}$ and $U_{\textrm{H}}$, we calculated a distance of $R= 2200$$^{+3800}_{-1200}$  pc for the outflow. Note that the uncertainties arise from the uncertainties in the measurements of $Q(\textrm{H})$, $n(\textrm{e})$, and $U(\textrm{H})$.
Similarly, we derive the top x-axis in Figure~\ref{figchi}. The \ion{S}{iv}'s upper limit on $n_{\textrm{e}}$ yields $R$>420 pc, consistent with the \ion{N}{iii}-based measurement.
\subsection{BH mass and the Eddington luminosity}

\begin{figure}
\includegraphics[width=\columnwidth]{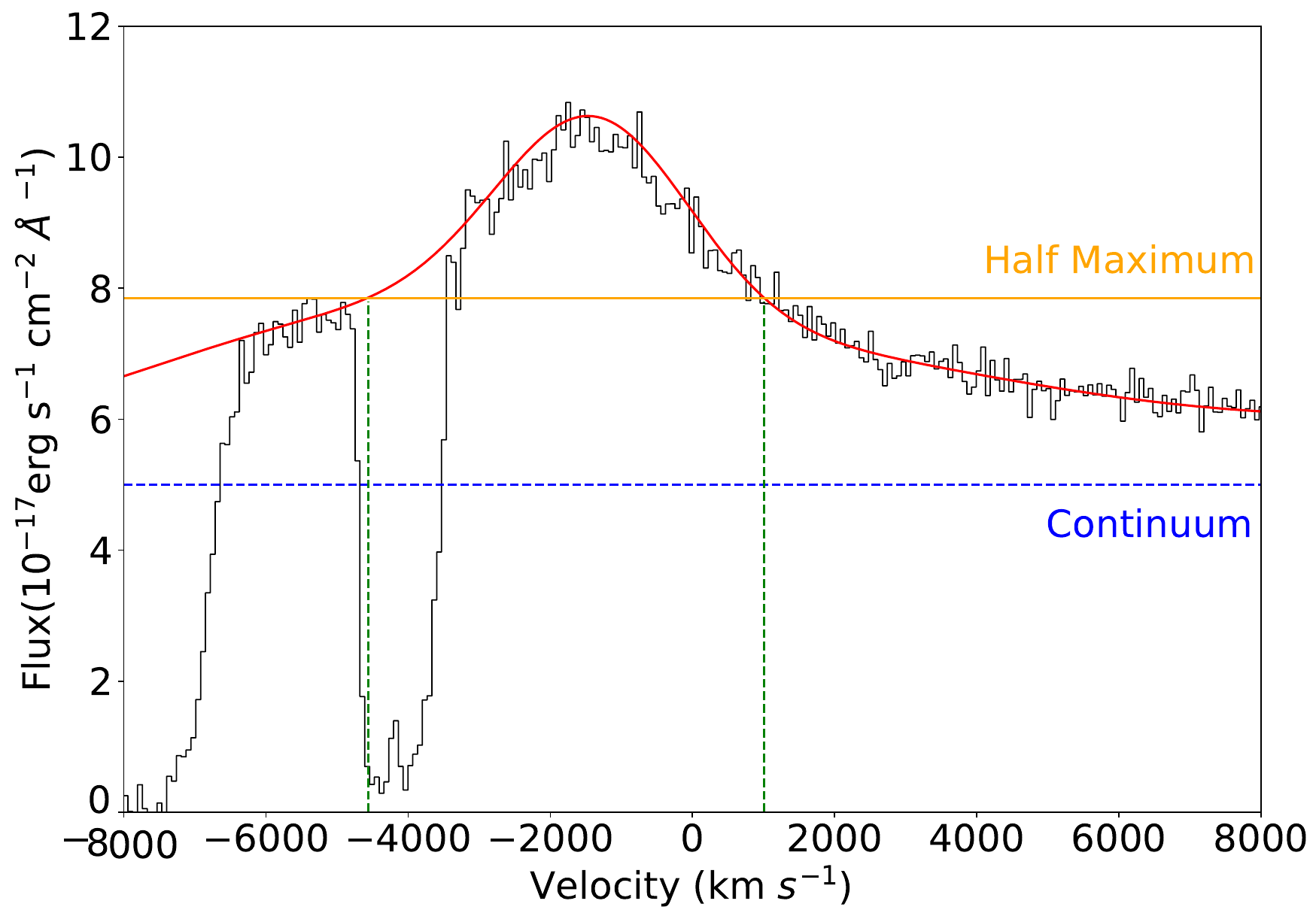}
\caption{ The spectrum in the region around the \ion{C}{iv} emission line. The red line shows our total emission model comprised of a continuum fit plus a model for the \ion{C}{iv} emission line.  The latter is comprised of two Gaussians, one narrow and one broad. The orange line indicates where the half maximum of the emission line model is, while the green dashed lines show the full width of the half maximum. The continuum is subtracted for FWHM determination. We determine a FWHM of 5600 km s$^{-1}$ and estimate that \ion{C}{iv} is blueshifted by about 1500 km s$^{-1}$.}
            \label{figC4}
\end{figure}

\cite{vest06} show that, assuming that the AGN broad emission line gas is virialized, one can use the \ion{C}{iv} emission line to estimate the BH's mass:
\begin{equation}
\log (M_{BH}(\textrm{\ion{C}{iv}}))=\log ( {[\frac{\textrm{FWHM(\ion{C}{iv})}}{1000  \textrm{kms}^{-1}}]^{2}[\frac{\lambda L_{\lambda}(1350\AA)}{10^{44} \textrm{ergs}^{-1}}]^{0.53}}) \label{eq2}
\end{equation}
However, the above equation does not consider the blue-shift in the emission line happening due to the presence of outflows or radiation pressure from the central source \citep[e.g.][]{coat16}. \cite{coat17} took this effects into account and revised equation~(\ref{eq2}) as below:
\begin{equation}
    \begin{aligned}
M_{BH}(\textrm{\ion{C}{iv}, Corr.})=10^{6.71}[\frac{\textrm{FWHM(\ion{C}{iv}, Corr.)}}{1000 \textrm{kms}^{-1}}]^{2}\times \\
[\frac{\lambda L_{\lambda}(1350\AA)}{10^{44} \textrm{ergs}^{-1}}]^{0.53}
    \end{aligned}\label{eq3}
\end{equation}
\noindent where:
\begin{equation}
    \begin{aligned}
\textrm{FWHM(\ion{C}{iv}, Corr.)}=\frac{\textrm{FWHM(\ion{C}{iv}, Meas.)}}{(0.41\pm0.02)\frac{\textrm{\ion{C}{iv} blueshift}}{1000 \textrm{kms}^{-1}}+(0.62\pm0.04)}
    \end{aligned} \label{eq4}
\end{equation}
Figure~\ref{figC4} displays the \ion{C}{iv} emission line region, along with our modeled emission line profile and continuum level. From this model, we estimate a FWHM of 5600 km~s$^{-1}$ and a \ion{C}{iv} blue-shift of almost 1500 km~s$^{-1}$. These calculations yield a BH mass of  $M$$_{\textrm{BH}}$=9.8$^{+1.2}_{-1.0}\times$10$^{8}$M$_{\astrosun}$
and 
$L$$_{\textrm{Edd}}$=1.1$^{+0.15}_{-0.13}\times$10$^{47}$(erg s$^{-1}$). 
\subsection{Outflow's energetics}
The mass-loss-rate ($\Dot{M}$) and kinetic luminosity ($\Dot{E}_{\textrm{K}}$) of the outflowing gas can be inferred using the following equations from \cite{borg12b}:
\begin{equation}
\Dot{M} \simeq 4\pi \Omega R N_\textrm{H} \mu m_{\textrm{p}} \nu \label{eq5}
\end{equation}
\begin{equation}
\Dot{E}_{\textrm{K}} \simeq \frac{1}{2}\Dot{M} v^{2}\label{eq6}
\end{equation}
\noindent where $\mu$ = 1.4 is the mean atomic
mass per proton, $v$ is the outflow's velocity, and m$_{p}$ is the proton's mass. $\Omega$ is the global covering factor defined as the fraction of the solid angle around 
the source covered by the outflow. 
There are several surveys in which \ion{C}{iv} BALs are detected in about 20$\%$ of all quasars (e.g., 
\citealt{hew03,dai12,gib09,all11}). This detection fraction is usually interpreted
to mean that all quasars have BAL outflows with $\Omega \approx$0.2, on average.
Using the derived values for $R$, $N_{\textrm{H}}$, and adopting an $\Omega$ of 0.2, for an outflow velocity of $-4300$~km~s$^{-1}$,  we calculate 
$\Dot{M}$=1070$^{+840}_{-700}$ M$_{\astrosun} \textrm{yr}^{-1}$ and $\log (\Dot{E_{\textrm{K}}}$)=45.8$^{+0.25}_{-0.46}$ [erg s$^{-1}$]. Note that the uncertainties in the kinetic luminosity are derived from the errors in the $\Dot{M}$ measurement. The errors on $\Dot{M}$ are estimated based on the elliptical shape of the photoionization solution (see figure.~\ref{NVUapp}), which indicates that the uncertainties in $N_{\textrm{H}}$ and $U_{\textrm{H}}$ are correlated \citep[see section 4.1 of ][for details on error propagation in the $\Dot{M}$ calculation]{walk22}.
Based on the values calculated for the bolometric and Eddington luminosities, we find that:
\begin{equation*}
\Dot{E_{\textrm{K}}}/{L_{\textrm{Bol}}} =5.2\% \hspace{1cm} \textrm{and}\hspace{1cm} \Dot{E_{\textrm{K}}}/{L_{\textrm{Edd}}} =5.7\%
\end{equation*}
An outflow with a $\Dot{E_{\textrm{K}}}/{L_{\textrm{Edd}}}$ of at least $\approx 0.5 \%$ is contributing significantly to the AGN feedback processes \citep{hop10}. 
The kinetic luminosity of the outflow in J1402 corresponds to more than 5$\%$ of the Eddington luminosity, suggesting a strong contribution to AGN feedback processes.

\section{Summary and Conclusions}
\label{sec:disc}
In this study, we investigate the properties of a high-energy outflow in  quasar J1402+2330, utilizing its spectrum from the DESI survey. While we detect two distinct absorption line outflows, the analysis centers on the low-velocity component due to its rich absorption features, which include ground and excited states from \ion{N}{iii}, \ion{S}{iv}, and \ion{C}{ii}. Using photoionization modeling and spectral diagnostics, we determine the physical characteristics of the outflow, including electron number density, hydrogen column density, and ionization parameter, which enable us to constrain the outflow’s distance from the central source to approximately 2200 parsecs.

Our results indicate that the outflow has a mass-loss rate on the order of a thousand solar masses per year and a kinetic luminosity that exceeds 5$\%$ of the Eddington luminosity. These findings imply that the outflow could play a significant role in AGN feedback, potentially influencing star formation and the interstellar medium within the host galaxy.  Table~\ref{tab3} provides a summary of all the physical properties calculated for this outflow system. This work demonstrates the utility of DESI data in resolving the detailed properties of quasar outflows and contributes to our understanding of the broader impact of AGN-driven feedback mechanisms.

\begin{table}[ht!]
\caption{\label{tab3}Properties of the mini-BAL outflow}
\centering
\begin{tabular}{lcc}
\hline
\\
$v_{\textrm{centroid}} $(km~s$^{-1}$)&&$-$4300 \\
\\
$\textrm{log}(N_\textrm{H})$[$\textrm{cm}^{-2}$]&&21.6$^{+0.5}_{-0.8}$ \\
\\
$\textrm{log}(U_\textrm{H})$&&$-1.1^{+0.5}_{-0.7}$ \\
\\
$\textrm{log}(n_\textrm{e})[\textrm{cm}^{-3}]$&&$2.85^{+0.20}_{-0.15}$ \\
\\
\textrm{R}(pc)&&$2200^{+3800}_{-1200}$ \\
\\
$\Dot{M}(\textrm{M}_{\astrosun}\textrm{yr}^{-1})$&&1070$^{+840}_{-700}$\\
\\
$\log (\Dot{E_{K}})$[erg s$^{-1}$]&&45.8$^{+0.25}_{-0.46}$ \\
\\
$\Dot{E_{K}}/{L_{Edd}}$($\%$)&&5.7$^{+5.8}_{-3.9}$\\
\\
\hline

\end{tabular}
\end{table}

\begin{acknowledgements}
We acknowledge support from NSF grant AST 2106249,
as well as NASA STScI grants AR-15786, AR-16600, AR-
16601 and AR-17556.This research uses services or data provided by the SPectra Analysis and Retrievable Catalog Lab (SPARCL) and the Astro Data Lab, which are both part of the Community Science and Data Center (CSDC) program at NSF's National Optical-Infrared Astronomy Research Laboratory. NOIRLab is operated by the Association of Universities for Research in Astronomy (AURA), Inc. under a cooperative agreement with the National Science Foundation.
\end{acknowledgements}

%
%

\end{document}